\documentclass[aps,prl,reprint,tightenlines,showpacs]{revtex4-1}

\usepackage{graphicx}

\usepackage{bm}

\begin{document}

\title{Superconducting inductive displacement detection of a microcantilever}

\author{A. Vinante}
\email{anvinante@fbk.eu}
\affiliation{Istituto di Fotonica e Nanotecnologie, CNR - Fondazione Bruno Kessler, I-38123 Povo, Trento, Italy.}

\date{\today}


\begin{abstract}
We demonstrate a superconducting inductive technique to measure the displacement of a micromechanical resonator. In our scheme, a type I superconducting microsphere is attached to the free end of a microcantilever and approached to the loop of a dc Superconducting Quantum Interference Device (SQUID) microsusceptometer. A local magnetic field as low as 100 $\mu$T, generated by a field coil concentric to the SQUID, enables detection of the cantilever thermomechanical noise at $4.2$ K. The magnetomechanical coupling and the magnetic spring are in good agreement with image method calculations assuming pure Meissner effect. These measurements are relevant to recent proposals of quantum magnetomechanics experiments based on levitating superconducting microparticles.
\end{abstract}

\maketitle

Several different techniques can be used to measure displacement of mechanical resonators. Because of simplicity and flexibility, capacitive and piezoelectric techniques are the most popular choice in Microelectromechanical systems (MEMS), while optomechanical techniques are usually the best choice when very high sensitivity is required. In this paper, we focus on the superconducting inductive transducer, which can be convenient in some special applications at cryogenic temperature. In this scheme, the distortion of a local magnetic field, caused by the motion of a superconductor, is measured by a nearby coil or a magnetic flux sensor like a Superconducting Quantum Interference Device (SQUID).

Superconducting inductive transducers at the macroscale with extremely low noise, of the order of $10^{-19}$ m/$\sqrt{\mathrm{Hz}}$, have been developed in the context of cryogenic bar detectors of gravitational waves \cite{allegro, gottardi}. Furthermore, they have been used for detecting the rotation of ultrasensitive space-based gyroscopes \cite{GPB} and for torsion balance experiments \cite{speake}. In contrast, superconducting inductive transducers at the microscale have received little or negligible consideration, although similar concepts have been widely exploited. Relevant examples are the magnetomotive detection of very high frequency resonators \cite{roukes}, the measurement of resonators embedded in SQUID structures \cite{etaki}, and the inductive detection of ferromagnetic particles \cite{usenko}.

Here, we demonstrate inductive detection of a superconducting microsphere attached to an oscillating microcantilever. In our scheme, shown in Fig. 1, the microsphere is approached to a loop of a gradiometric SQUID microsusceptometer, which comprises a field coil loop concentric to the SQUID loop. When a magnetic field is locally generated by applying a current to the field coil, a position-dependent distribution of Meissner shielding currents is developed in the microsphere, coupling a magnetic flux into the SQUID. A displacement of the microsphere due to cantilever motion will therefore produce a measurable change of the magnetic flux threading the SQUID.
\begin{figure}[!ht]
\includegraphics{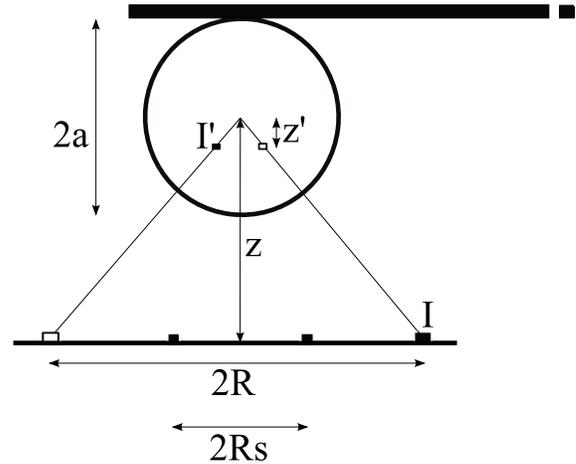}
\caption{Simplified scheme and model of the experiment. A type I superconducting microsphere of radius $a$ is glued on the end of a microcantilever and positioned at a distance $z$ from a SQUID loop (inner loop with radius $R_S$) and a field coil (outer loop concentric to SQUID loop with radius $R$). When a current $I$ is applied to the field loop, the Meissner response of the superconductor is equivalent, outside the superconductor, to that of an image current $I^{\prime}$ at a distance $z^{\prime}$ from the sphere center. Hollow and solid rectangles represent inward and outward current. The $z$-dependent interaction force $F \left( z \right)$ between $I$ and $I^{\prime}$ induces a magnetic spring for cantilever motion along $z$. Similarly, Detection of cantilever motion is enabled by the $z$-dependent magnetic flux $\Phi \left( z \right)$ coupled by the image current into the SQUID loop.}  \label{schema}
\end{figure}

If the superconductor is of type I, pure Meissner effect is assumed to take place. As the magnetic field applied by the local field coil is nonuniform, modeling of the Meissner response is in general non-trivial. However, assuming (i) spherical shape of the particle and (ii) motion of the particle center along the symmetry axis of the field coil, a convenient solution based on the image method is found \cite{lin}. In particular, the Meissner response outside the sphere is equivalent to that of an image current $I^{\prime}=-\left( b/a \right) I$ placed at a distance $z^{ \prime} = \left( a^2/b^2 \right) z$ from the sphere center (Fig. 1), where $I$ is the current in the field coil, $a$ is the particle radius and $b=\sqrt{R^2+z^2}$, with $R$ coil radius and $z$ height of the sphere center above the coil. The interaction of the image current with the coil current produces a repulsive levitating force:
\begin{widetext}
\begin{equation}
F\left( z \right) = \mu _0 I^2 R^2 az\left( {R^2  - a^2  + z^2 } \right)\int\limits_0^\pi  {\frac{{\cos \theta d\theta }}{{\left[ {a^4  + \left( {R^2  + z^2 } \right)^2  - 2a^2 \left( {R^2  + z^2 \cos \theta } \right)} \right]^{\frac{3}{2}} }}} ,   \label{flev}
\end{equation}
\end{widetext}
and a magnetic spring $k_m = - dF/dz$.
Similarly, using elementary formulas for the mutual inductance between coaxial circular loops, one can derive the flux $\Phi \left( z \right)$ coupled into the SQUID by the image current, and the magnetomechanical coupling $\Phi_z=d\Phi/dz$, which can be expressed again in terms of elliptic integrals.

In our experiment, the superconducting particle is picked from $99.5\%$ pure Pb powder (Goodfellow) and is approximately spherical, with radius $a=\left( 11 \pm 1 \right)$ $\mu$m. The particle is epoxy-glued on a tipless AFM cantilever with dimensions $450 \times 56 \times 2.2$ $\mu$m$^3$ and nominal spring constant $k=\left( 0.28 \pm 0.02 \right)$ N/m. The SQUID susceptometer \cite{vinante3} is composed of two identical distant loops each with radius $R_S=10$ $\mu$m. The field coil is composed of two identical loops concentric to each SQUID loop, with radius $R=24$ $\mu$m. The field coil loops are connected in a gradiometric configuration, so that the direct coupling of one loop to the SQUID is canceled by the coupling of the other loop within one part per $600$. The cantilever is manually aligned above the SQUID with the help of a macor spacer and held in position by a brass spring. The position of the sphere is estimated by optical microscope inspection as $\left( x_0,y_0,z_0 \right) =\left( 0 \pm 5,0 \pm 5,40 \pm5 \right)$ $\mu$m, with respect to the center of the SQUID loop, and the motion corresponding to the fundamental flexural mode is thus approximately along the $z$ axis. The system is enclosed in a shielded copper box and cooled in a vacuum-tight probe immersed in liquid helium. Pressure is measured by a Penning gauge placed at room temperature.

Dc currents $I$ up to $30$ mA can be applied to the field-coil under stable conditions. Higher currents lead to a sudden increase in the power dissipated and therefore to a thermal drift, probably caused by the superconducting to normal transition of the field coil. The maximum dc current $I=30$ mA corresponds to a field $B\simeq 100$ $\mu$T and field gradient $\partial B/\partial z \simeq 5$ T/m in the sphere center.
When a dc bias current $I$ is applied, the cantilever fundamental mode can be excited by superimposing a small ac current $I_{ac}$ at the resonant frequency. This causes, according to Eq. (\ref{flev}), a modulation of the force given by $F_I I_{ac}$, where $F_I=\partial F/\partial I$. After switching off the ac excitation, the cantilever ring-down is measured by a lock-in amplifier and the free resonant frequency $f_0$ and the quality factor $Q$ can be estimated with high precision.

The resonant frequency at very low dc current is $f_0 = 8318.685$ Hz. A non-zero current $I$ generates a magnetic spring that shifts the frequency up. Fig. 2 shows the measured resonant frequency shift $\delta f_0$ as a function of the applied current $I$. The purely quadratic dependence expected from Eq. (\ref{flev}) is observed. The quadratic fit with $\delta f_0=\alpha I^2$ yields $\alpha=\left(  15.1 \pm 0.1 \right)$ Hz/A$^2$.
The quality factor is slightly dependent on the pressure $P$ of the residual gas. The best $Q$ factor measured at a nominal pressure $P=1.3\times10^{-5}$ mbar is $Q=6.5 \times 10^5$. For a given pressure $P$, $Q$ is independent of the current $I$ within the measurement repeatibility of the order of $0.5$ percent.
\begin{figure}[!ht]
\includegraphics{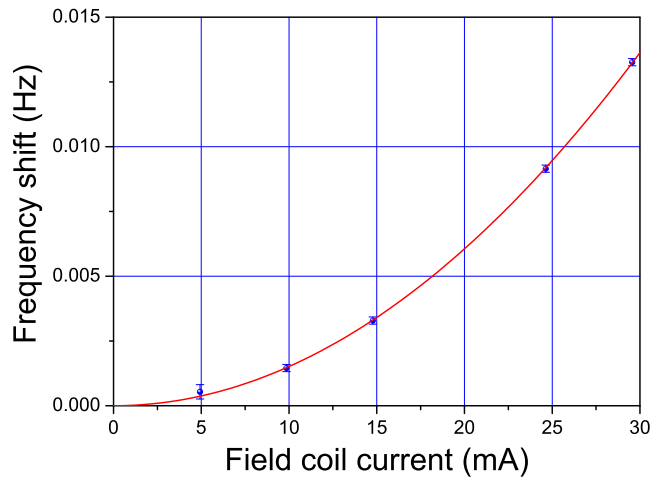}
\caption{(Color Online.) Frequency shift of the cantilever fundamental mode resonance, caused by the magnetic spring, as function of the current $I$ applied to the field coil. The curve is a pure quadratic fit with the function $\delta f_0 = \alpha I^2$ .}  \label{alpha}
\end{figure}

A direct absolute measurement of the magnetomechanical coupling $\Phi_z$ is not possible with our setup. However, we can measure the frequency response of the signal measured by the SQUID when an ac current excitation in the field coil is swept through resonance. The ratio of the measured SQUID flux $\Phi$ to the excitation current $I_{ac}$ around resonance is given by the following expression:
\begin{equation}
T\left( \omega  \right) = \frac{{\Phi }}{{I_{ac} }} = \Phi _I^{\left( {dir} \right)}  + \frac{{F_I \Phi _z }}{k}\frac{{\omega _0 ^2 }}{{\left( { - \omega ^2  + \omega _0 ^2 } \right) + i\left( {\frac{{\omega \omega _0 }}{Q}} \right)}} .  \label{TF}
\end{equation}
Here, $\Phi _I^{\left( {dir} \right)}$ is a frequency-independent term expressing the direct cross-talk between field coil and SQUID.
\begin{figure}[!ht]
\includegraphics{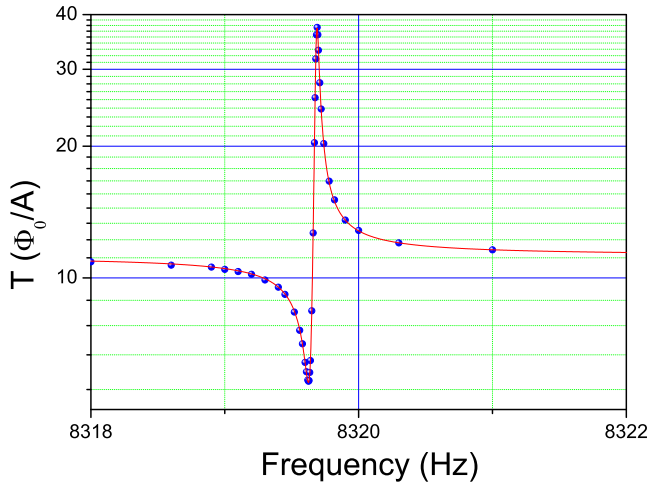}
\caption{(Color Online.) Resonant response of the cantilever around resonance for $I=29.5$ mA, expressed as ratio between the flux $\Phi$ measured by the SQUID and the ac current $I_{\mathrm{ac}}$ superimposed to the bias dc component $I$. The solid curve represents the best fit with the module of Eq. (\ref{TF}).}  \label{B}
\end{figure}

The experimental measurement of the module of the transfer function $T\left( \omega  \right)$ is shown in Fig. 3, and is performed at a pressure $P=5.6 \times 10^{-4}$ mbar and dc bias current $I=29.5$ mA. The data are well fitted by the module of Eq. (\ref{TF}). From the fitting curve parameters we can extract the combined product $\beta=F_I \Phi_z Q/k=\left( 31.6 \pm 0.1 \right) $ $\Phi_0$/A, which is expressed geometrically by the difference between the maximum and minimum of the curve. The quality factor extracted by the fit $Q=\left( 2.58 \pm 0.01 \right) \times 10^5$ is consistent with the value estimated by the ringdown method at this pressure.

As last measurement, we have estimated the cantilever narrow-band noise at the same pressure of the transfer function measurement. $900$ datapoints were acquired by a narrow-band lock-in amplifier referenced at $f_0$, with a sampling time of $16$ s, larger than the resonator time constant in order to get uncorrelated samples.  The experimental histogram of the lock-in measured energy, in units of squared flux in the SQUID, is shown in Fig. 4 and is exponentially distributed. From the slope of the exponential fit we can estimate the mean energy, upon subtraction of the contribution of the SQUID wideband noise in the lock-in bandwidth. The corrected mean value of the cantilever noise is $\left\langle {\Phi_n ^2 } \right\rangle = \left( 1.5 \pm 0.2 \right) \times 10^{-6}$ $\Phi_0^2$.
\begin{figure}[!ht]
\includegraphics{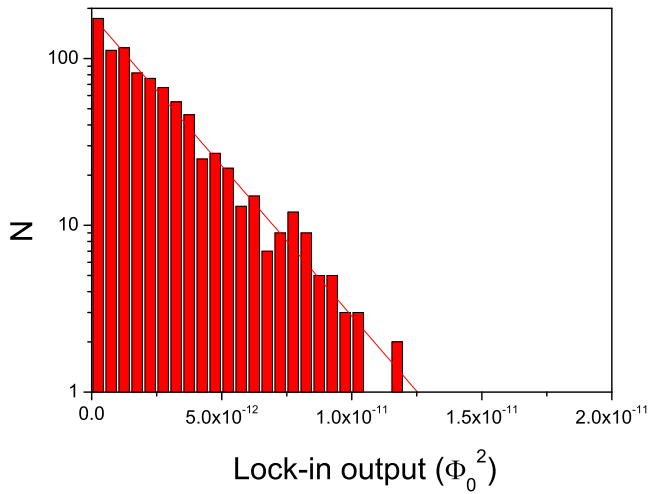}
\caption{(Color Online.) Histogram of the energy signal of a lock-in centered on the resonant frequency of the cantilever. The data are well fitted by an exponential function, consistently with a thermal distribution. From the fitting slope, and upon subtracting the SQUID wideband noise contribution, we can estimate the mean energy $\left\langle {\Phi_n ^2 } \right\rangle $ of the cantilever, in unit of $\Phi_0^2$ at the SQUID.}  \label{thermal}
\end{figure}

We can now compare the three independent measurements of the magnetic spring coefficient $\alpha$, the magnetomechanical coupling product $\beta$, and the resonator narrowband noise $\left\langle {\Phi_n ^2 } \right\rangle $ with the theoretical predictions of the image current model.
To simplify the analysis, we assume that the center of mass of the microsphere is moving exactly along the SQUID symmetry axis $z$ (ie $x_0,y_0=0$), and leave only the position $z_0$ along the $z$ axis as free parameter. Under these assumptions, we find that the measured values of $\alpha$ and $\beta$ are simultaneously reproduced for $z_0=36.5$ $\mu$m, which is consistent with the height estimated at room temperature by optical microscope inspection. The values predicted by the model are $\alpha=\left( 15 \pm 1 \right)$ Hz/A$^2$ and $\beta=\left( 31 \pm 6 \right) $ $\Phi_0$/A. The latter is calculated for the experimental conditions of the transfer function measurement, $I=29.5$ mA and $Q=2.58 \times 10^5$. The main contribution to the error bar comes from the uncertainty on the spring constant $k$, which is calculated from the geometrical dimensions of the cantilever.

We can use the same model to predict the thermomechanical noise coupled into the SQUID. Assuming a temperature $T=4.2$ K, the thermomechanical noise coupled into the SQUID is calculated, using the parameters estimated above, $\left\langle {\Phi_n ^2 } \right\rangle = \left( 1.2 \pm 0.3 \right) \times 10^{-12}$ $\Phi_0^2$, which is consistent with the measured value within the error bar. This shows that the cantilever motion is essentially thermally-limited, and that the sensitivity of the SQUID-based inductive detection technique is sufficient to measure the thermomechanical noise.

Considering the possible systematic errors that may affect the experimental estimation of the relevant parameters of the model, the agreement between theory and experiment is excellent. We conclude that the model, based on pure Meissner effect and image method, provides an adequate description of the system. From the experimentally-validated model, we can further estimate that the SQUID wideband noise floor corresponds to an equivalent displacement noise $S_z = \left( 20 \pm 3 \right)$ pm/$\sqrt{\mathrm{Hz}}$, for $I=29.5$ mA.

The demonstrated technique is relevant to quantum magnetomechanics experiments, recently proposed in Ref. \cite{romero}, which could pave the way to macroscopic tests of quantum mechanics and unconventional collapse models \cite{romero2}. These proposals are based on Meissner levitation of superconducting microparticles, with an inductive coupling to a superconducting detection loop, like a SQUID or a qubit. Our work demonstrates that the theoretical predictions based on pure Meissner effect are reliable for type I superconducting microparticles, in particular made of lead, close to a current loop \cite{lin}. It remains a challenge to demonstrate stable levitation in a zero-field trap of a superconducting particle, not attached to a cantilever, and to achieve the ultrahigh mechanical quality factor suggested in Ref. \cite{romero}.

The author acknowledges the support of the RESTATE Programme, co-funded by the European Union under the FP7 Marie Curie Action COFUND - Grant agreement n. 267224.

\end{document}